\documentclass[aps,prl,twocolumn,groupedaddress,longbibliography,nofootinbib]{revtex4-2}
\usepackage{amsmath}
\usepackage{amssymb}
\usepackage{graphicx}
\usepackage{xcolor}

\begin{document}
\title{Linear-response magnetoresistance effects in chiral systems}
\author{Xu~Yang}
\email[]{xu.yang@rug.nl}
\affiliation{Zernike Institute for Advanced Materials, University of Groningen, NL-9747AG Groningen, The Netherlands}
\author{Bart~J.~van~Wees}
\affiliation{Zernike Institute for Advanced Materials, University of Groningen, NL-9747AG Groningen, The Netherlands}
\date{\today}
\begin{abstract}
The chirality-induced spin selectivity (CISS) effect enables the detection of chirality as electrical charge signals. It is often studied using a two-terminal circuit geometry where a ferromagnet is connected to a chiral component, and a change of electrical resistance is reported upon magnetization reversal. This is however not expected in the linear response regime because of compensating reciprocal processes, limiting the interpretation of experimental results. Here we show that magnetoresistance effects can indeed appear even in the linear response regime, either by changing the magnitude or the direction of the magnetization or an applied magnetic field. We illustrate this in a spin-valve device and in a chiral thin film as the CISS-induced Hanle magnetoresistance (CHMR) effect. \textcolor{black}{This effect helps to distinguish spin-transport-related effects from other effects, and can thereby provide further insight into the origin of CISS.}
\end{abstract}
\maketitle

Chirality-induced spin selectivity (CISS) describes the spin-dependent electron transport through a chiral (molecular) system~\cite{naaman2019chiral}. It promises novel applications such as using chiral materials for information technologies (spintronics)~\cite{michaeli2017new} and using spintronic techniques for chemistry and biology~\cite{banerjee2018separation}. \textcolor{black}{The understanding of CISS has been greatly advanced by extensive experimental and theoretical research~\cite{ray1999asymmetric,gohler2011spin,xie2011spin,suda2019light,inui2020chirality,liu2020linear,kim2021chiral,shiota2021chirality,guo2012spin,medina2012chiral,medina2015continuum,varela2016effective,michaeli2015origin,geyer2019chirality,fransson2019chirality,matityahu2016spin,dalum2019theory,nurenberg2019evaluation,zollner2019chiral,du2020vibration,zhang2020chiral,utsumi2020spin,fransson2020vibrational,fransson2021charge,liu2021chirality}, but there remains a major question: how can it be detected as an electrical signal in the linear response regime, especially when the chiral system is coupled to a ferromagnet~\cite{yang2019spin,yang2020detecting,evers2021theory}?}

\textcolor{black}{Electron transport in the linear response regime is subject to fundamental laws of thermodynamics, in particular, the Onsager reciprocity~\cite{onsager1931reciprocal_i} and the subsequently formulated B{\"u}ttiker reciprocity theorem~\cite{buttiker1985generalized,buttiker1988symmetry}. For CISS, which we define as the spin--charge current interconversion within a chiral system, reciprocity implies strict symmetry relations between the spin-to-charge and charge-to-spin conversion mechanisms. Therefore, theoretical considerations must include both mechanisms in a self-consistent way. This becomes extra important when the chiral system is coupled to a ferromagnet since they both provide spin-to-charge and charge-to-spin conversion. Here a self-consistent treatment predicts a strict zero result in the linear response regime for a two-terminal circuit containing a chiral system and a ferromagnet (e.g. a conducting AFM geometry for chiral molecules using a magnetic substrate or a magnetic probe, or a spin-valve device geometry with ferromagnetic and chiral molecular layers): There is no change of electrical resistance upon full magnetization reversal, i.e. no magnetoresistance (MR), as was previously discussed~\cite{yang2019spin,yang2020detecting,evers2021theory}. Building on this, we predicted that any possible (magnetization-reversal-induced) MR beyond linear response must first appear linear in bias, just as in the electrical magneto-chiral effect~\cite{yang2020detecting,rikken2001electrical}. These results were supported by further theoretical analyses~\cite{utsumi2020spin,liu2021chirality}, but still intrigued much discussion since in experiments where a region of linear current--voltage behavior can be identified, many do in fact observe different slopes for opposite magnetization directions, implying a nonzero MR in the linear response regime~\cite{liu2020linear,mishra2020spin,lu2021spin,kim2021chiral,mondal2021spin,lu2021spin,mondal2015chiral,naaman2020comment,yang2020reply}.} 

\textcolor{black}{Note that this discussion focuses on the MR generated by the entire two-terminal circuit, but NOT the spin (current) polarization generated by the chiral system. These two concepts are fundamentally different. The MR generated by the entire circuit is a magnetic-field-dependent electrical charge signal. In contrast, the spin (current) polarization generated by the chiral system is not a charge signal and cannot be directly measured. The Onsager reciprocity implies that the MR related to spin-transport must vanish in the linear response regime. This does not impose constraints on MR of other origin or on the spin (current) polarization generated by the chiral system. This major distinction between MR and spin (current) polarization was previously addressed~\cite{yang2020reply,yang2020detecting,yang2019spin,fransson2019chirality,liu2021chirality}, but is sometimes still overlooked~\cite{waldeck2021spin,varela2021response}.} 

\textcolor{black}{We emphasize that the Onsager reciprocity is based on the fundamental microscopic reversibility and therefore should hold universally. This was repeatedly demonstrated in experiments~\cite{leturcq2006magnetic,peled2004symmetries}, including for a spin--orbit system coupled to a ferromagnet~\cite{lee2018multi} and for strongly correlated systems~\cite{puller2009breaking}. Any deviation from the Onsager reciprocity indicates the possible presence of effects other than charge and spin transport. These yet-unknown effects may as well depend on chirality and magnetic field/magnetization, and may give rise to MR. It is therefore extremely important to distinguish spin-transport-related effects from other effects.}

\textcolor{black}{Important to realize, the zero (transport-induced) MR predicted by Onsager reciprocity requires the full reversal of magnetization or magnetic field.} This leaves room for linear-response MR effects to arise by varying the magnitude or direction of the magnetization or magnetic field. We illustrate these new MR effects and show how they can be detected in common device geometries, such as in a spin valve and in a chiral two-dimensional (2D) film.

\section{Magnetoresistance in two-terminal spin valve geometry}
\begin{figure}[ht!]
	\includegraphics[width=\linewidth]{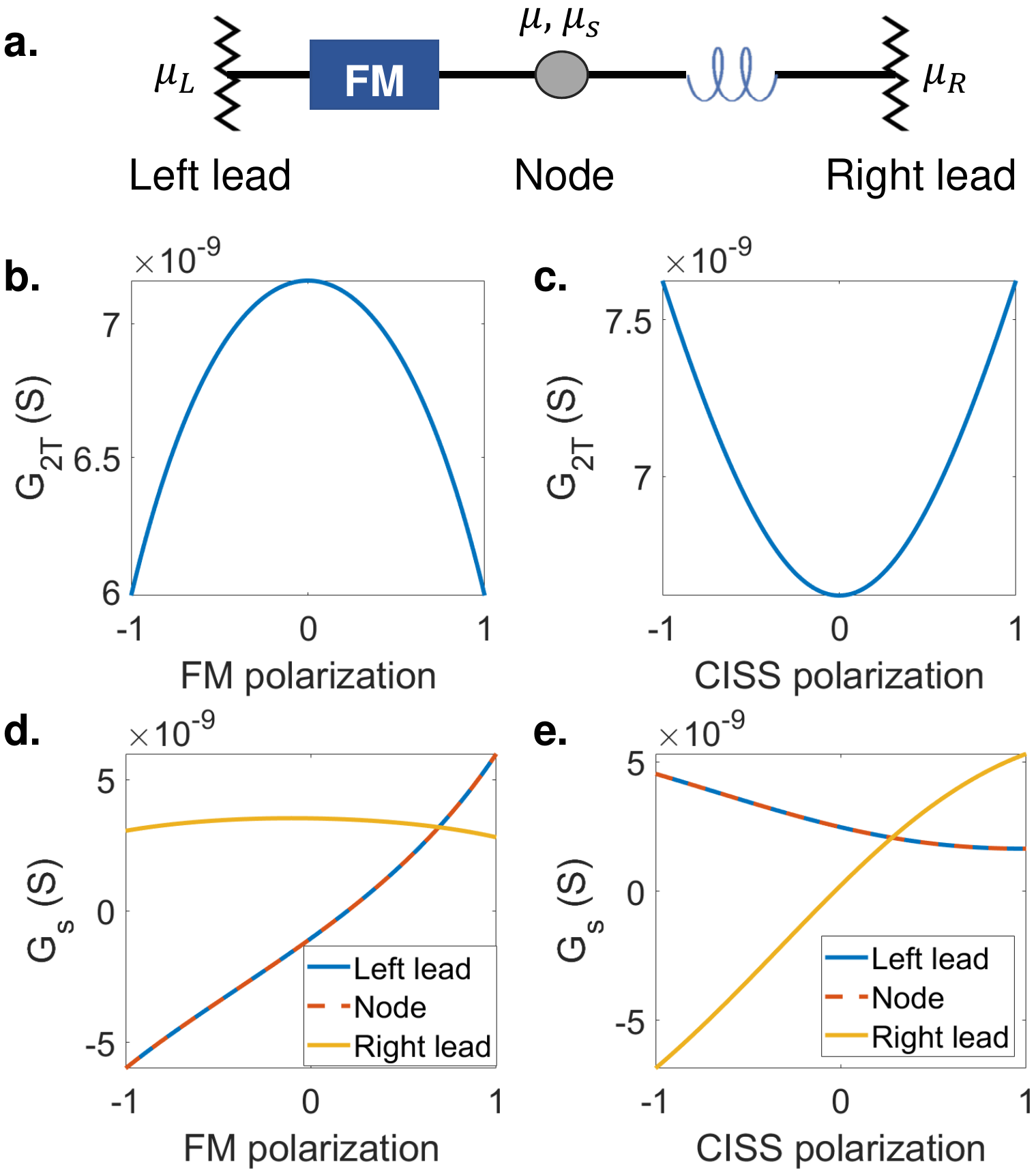}
	\caption{\label{fig:2TMR}The effect of ferromagnet and CISS polarization on the charge and spin currents in a two-terminal spin valve geometry. \textbf{a.} The two-terminal spin valve device contains one ferromagnet and one chiral component (blue helix) connected by a node (gray circle) between two electrodes (wavy vertical lines). The chiral component is assumed to display CISS with a set of spin-dependent electron transmission probabilities and the accompanying self-consistent reflection probabilities. The node characterizes momentum relaxation processes and is described by an electrochemical potential $\mu$ and a spin accumulation $\mu_s$. The electrodes are characterized by electrochemical potentials $\mu_L$ and $\mu_R$, which are supplied by a bias voltage $V=-(\mu_L-\mu_R)/e$ ($e$ is elemental charge), which drives the charge and spin currents. \textbf{b-c} The two-terminal (charge current) conductance $G_{2T}$ as a function of FM polarization and the CISS polarization (see text for definition). \textbf{d-e} The spin current conductance $G_s$ (spin current per unit two-terminal voltage) in the left lead, the node, and the right lead, as a function of the same variables. The blue solid and the red dashed lines overlap because of the identical spin currents in the left lead and the node.}
\end{figure}

We first show how an magnetoresistance (MR) can arise in the linear response regime in the two-terminal spin valve geometry introduced in Refs.~\cite{yang2020detecting,yang2019spin} (see Fig.~\ref{fig:2TMR}\textbf{a}). The circuit contains a single ferromagnet and a chiral component that displays CISS in the linear response regime. The two spin--charge interconverting elements are connected by a node, which models relaxation processes within the device. We do not assume a specific microscopic mechanism for CISS, but only phenomenologically characterize the spin (current) polarization generated by a charge current through the chiral component using a \textit{CISS polarization} parameter. \textcolor{black}{(Note that this is the $P_t$ parameter in the coupled charge--spin transport matrix $\mathcal{T}$ introduced in Ref.~\cite{yang2020detecting}, also see Appendix~A, Eq.~2. Below, when evaluating the effect of changing $P_t$, the relevant parameters of matrix $\mathcal{T}$ are adjusted self-consistently to take into account both of the reciprocal spin--charge interconversion mechanisms.)} Similarly, the ferromagnet is characterized by the \textit{FM polarization} parameter, which describes the spin polarization of the electrons transmitted through the ferromagnet (tunnel junction).

We use the transport matrix formalism introduced in Ref.~\cite{yang2020detecting} to calculate both charge and spin currents driven by a two-terminal voltage, and show here the corresponding linear-response conductances as a function of the two polarization parameters. As plotted in Fig.~\ref{fig:2TMR}\textbf{b-c}, the two-terminal (charge current) conductance $G_{2T}$ depends evenly on the FM polarization and the CISS polarization. This confirms that neither magnetization reversal (sign change for FM polarization) nor chirality reversal (sign change for CISS polarization) can change the two-terminal charge current and generate an MR signal. This vanishing MR is the consequence of the exact compensation of the spin-to-charge and charge-to-spin conversion mechanisms, which is inherent to the fundamental nature of linear response. It does not depend on microscopic details of the device, such as the exact conduction mechanism, the electrode material, and/or the presence of spin--orbit interaction. This was first established by Onsager as a fundamental law of thermodynamic processes~\cite{onsager1931reciprocal_i}, and then elaborated by B{\"u}ttiker for electron transport systems like the one discussed here~\cite{buttiker1985generalized,buttiker1988symmetry}.

Nevertheless, the dependence suggests that changing the magnitude of the FM polarization or the CISS polarization can indeed change the two-terminal conductance and give rise to an MR-type signal. In practice, the FM polarization can be experimentally tuned by rotating the magnetization direction using an external magnetic field, which changes the net magnetization projected along the relevant direction (collinear to the charge current). The CISS polarization, on the other hand, is in most cases fully determined by the choice of material and is not easily accessible as an experimental variable, but it can be tunable for some chiral molecules, such as a molecular motor that gradually switches (sign of) chirality under light illumination~\cite{koumura1999light,suda2019light}.

We also show in Fig.~\ref{fig:2TMR}\textbf{d-e} the linear response spin current per unit two-terminal voltage, which we define as the spin current conductance $G_s$. It is evaluated separately in the left lead, the node, and the right lead, and we show here its dependence on the FM polarization and the CISS polarization. An important observation here is that the spin current is not conserved. It is the same in the left lead and in the node since the ferromagnet (ideally) does not introduce spin relaxations mechanisms, but it is different in the right lead because of the spin-flip mechanisms that are inherent to CISS-type spin-dependent transmission~\cite{yang2019spin,yang2020detecting,bardarson2008proof,matityahu2016spin,utsumi2020spin}. The spin current in the left lead and the node depends more strongly on the FM polarization, while in the right lead it depends more strongly on the CISS polarization. \textcolor{black}{Note that even when the CISS polarization is zero, the spin-flip mechanisms are still present, and that is why in Fig.~\ref{fig:2TMR}\textbf{e}, $G_s$ is different for the node and the right lead at zero CISS polarization.}

Another general observation is that the spin currents are not even-functions of the polarization parameters. Therefore, a magnetization reversal or chirality reversal will change the spin currents (and spin polarization) in the leads and the node. This is indeed in agreement with theories that address the spin polarization in such device geometries~\cite{medina2015continuum,medina2012chiral,guo2012spin,varela2016effective,michaeli2015origin,zollner2019chiral,fransson2019chirality,geyer2019chirality,du2020vibration,zhang2020chiral}. Important to realize, this changing spin polarization cannot be experimentally detected as a charge signal because of the zero MR upon polarization reversal.

\textcolor{black}{The implication of reciprocity goes beyond transport driven by a charge voltage. In Appendix~C, we illustrate how spin accumulation in the leads drives spin transport and how it gives rise to specific symmetry relations between spin currents in the leads.}

Note that here we have omitted interface effects such as spin-mixing conductance~\cite{brataas1999spin} and torque-induced spin precession~\cite{adagideli2006detection}, which may also apply to CISS. These effects \textcolor{black}{may generate spin components that are orthogonal to the spin orientation of interest, thereby reducing the net spin density for that orientation and affecting our results quantitatively, but not qualitatively.}

\section{Magnetoresistance in 2D chiral film}
Next, we introduce an alternative approach to detect CISS in linear response. We extend our description of the chiral elements to model a 2D chiral film. We illustrate how spin precession dynamics in such a film can give rise to MR signals that depend on both the orientation and the magnitude of an applied magnetic field.

\subsection{Network model for 2D chiral film}

\begin{figure*}[htb!]
	\includegraphics[width=\linewidth]{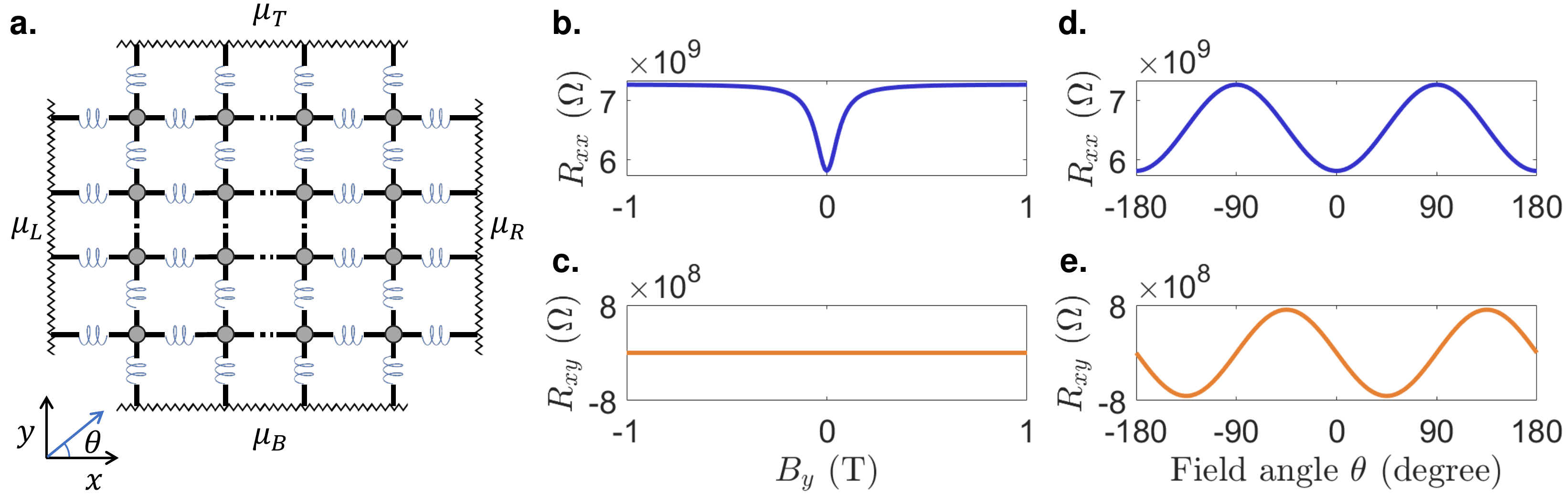}
	\caption{\label{fig:model}Network model and the CISS-induced Hanle Magnetoresistance (CHMR) effect. \textbf{a.} A network model for a uniform 2D chiral film. It constitutes a number of chiral elements (blue helical wires) connected to each other via nodes (gray circles). The four edges are electrically contacted by nonmagnetic electrodes (wavy lines) characterized by their electrochemical potentials $\mu_L$ (left), $\mu_R$ (right), $\mu_T$ (top), and $\mu_B$ (bottom), respectively. \textbf{b-c.} The longitudinal ($R_{xx}$) and transverse ($R_{xy}$) sheet resistances for a charge current along the $x$-direction, as a function of a perpendicular magnetic field strength $B_y$. \textbf{d-e.} $R_{xx}$ and $R_{xy}$ under an in-plane rotating magnetic field (at constant strength), plotted as a function of its angular direction $\theta$ with respect to the $x$-axis.}
\end{figure*} 

Device-based CISS experiments often employ chiral thin films, where the spin--charge interconversion can be described using spin currents and spin accumulations in the film. Such films often consist of unaligned chiral molecules, and it is their averaged property that is experimentally accessible. For a qualitative analysis, we can consider a uniform film and model it using a 2D network of (identical) chiral elements connected via (identical) nodes, as illustrated in Fig.~\ref{fig:model}\textbf{a}. With this, we can separately treat the effects due to CISS (spin--charge conversion and spin-flip reflection) and those generally present in a device (additional spin relaxation and precession). The nodes also allow us to evaluate voltages. This treatment is for the ease of modeling and does not imply using physically distinct materials within an actual film.

We consider the coupled charge and spin transport associated with an $x$-direction charge current $I_{x}$ and the accompanying spin currents within the film. These combined give rise to a longitudinal charge voltage $V_{xx}=-(\mu_L-\mu_R)/e$ (between the left and right electrodes) and a transverse charge voltage $V_{xy}=-(\mu_T-\mu_B)/e$ (between the top and bottom electrodes). This is characterized by a longitudinal sheet resistance $R_{xx}=V_{xx}/I_x$ and a transverse sheet resistance $R_{xy}=V_{xy}/I_x$, which are experimentally accessible. Here we assumed the film is a square, and the effect of film geometry is discussed in the Appendix~A. Because of the coupled charge and spin transport due to CISS, these resistances depend on spin dynamics such as relaxation and precession around an external magnetic field, which can be quantitatively evaluated using the 2D network model. 

We assume the spin accumulation is evenly distributed in the nodes as an approximation for the averaged property of the film. Each node (or per unit area) supports a spin accumulation $\boldsymbol{\mu_s}=(\mu_{sx}, \mu_{sy}, \mu_{sz})=\boldsymbol{n_s}/2\nu$, where $\boldsymbol{n_s}=(n_{sx}, n_{sy}, n_{sz})$ is the spin density (with $x$, $y$, and $z$ labeling the spin orientation), and $\nu$ is the averaged density of states in the film. These spins are injected in the nodes and they undergo other dynamical processes. Here we consider the effect of spin relaxation (characterized by a spin lifetime $\tau_s$) and Larmor spin precession (induced by a magnetic field $\boldsymbol{B}$). Note that we assume the spin lifetime $\tau_s$ and the response of spins to the magnetic field are isotropic (in spin space).

We are interested in the steady-state condition of these dynamical processes under the presence of the charge current $I_x$. The spin injection mechanism that we focus on here is the charge to spin conversion due to CISS, and we write the total injected spin current in a node as $\boldsymbol{I_{s,CISS}}$ (see Appendix~A for detail). Note that there are also other spin injection mechanisms such as (optical or radio-frequency) spin pumping, but these are beyond the scope of this discussion. The steady-state condition is
\begin{equation}\label{eqn:steady}
	0=\frac{\partial\boldsymbol{n_s}}{\partial t}=-\frac{\boldsymbol{I_{s,CISS}}}{e}+\frac{2\mu_B}{\hbar}\boldsymbol{B}\times \boldsymbol{n_s}-\frac{\boldsymbol{n_s}}{\tau_s},
\end{equation}
where $\mu_B$ is the Bohr magneton, and $\hbar$ is the reduced Planck's constant. 

The key to coupling this spin dynamics to charge signals is the spin--charge interconversion. Via CISS, the charge current $I_x$ generates a collinear spin current $I_{s,CISS,x}$ and spin density $n_{sx}$. These spins precess due to the (noncollinear) magnetic field, which generates spin currents and spin densities along $x$, $y$, and $z$ (in spin space). These are then converted to charge currents along $x$, $y$, and $z$ directions due to the reciprocal effect of CISS, thereby affecting the charge signal in both longitudinal and transverse directions.

\subsection{CISS-induced Hanle Magnetoresistance (CHMR)}
We analytically solve the expression of the longitudinal and transverse sheet resistances $R_{xx}$ and $R_{xy}$, and \textcolor{black}{the result} demonstrates two ways of detecting the linear-response MR due to the spin precession dynamics. In Fig.~\ref{fig:model}\textbf{b-c}, we consider a magnetic field along the $y$-direction and show the effect of varying its strength. The longitudinal sheet resistance $R_{xx}$ depends on the field strength but not its sign. It reaches a minimum at zero field and increases with increasing field strength. \textcolor{black}{This is because the finite perpendicular magnetic field induces spin precession, which reduces the net spin density along the direction that was originally aligned to be converted to a longitudinal charge current via inverse CISS. This precession-induced spin density reduction reduces the longitudinal charge current by spin--charge interconversion, and thereby exhibits as an increased $R_{xx}$.} Meanwhile, the transverse sheet resistance $R_{xy}$ remains zero, because the precessing spins are always orthogonal to $y$ and thereby cannot give rise to charge transport along the $y$-direction.

In Fig.~\ref{fig:model}\textbf{d-e}, we consider a magnetic field of 1~T rotating within the sample plane, and plot the sheet resistances as a function of its angle $\theta$ with respect to $x$. Both $R_{xx}$ and $R_{xy}$ are sinusoidal/cosinusoidal to the field angle with a period of $180^{\circ}$ but with a angular phase shift of $45^{\circ}$. The $R_{xx}$ is at its minimum when the field is collinear with the charge current and at its maximum when they are orthogonal, and at these two directions the $R_{xy}$ is zero. The range of $R_{xx}$ is the same in Fig.~\ref{fig:model}\textbf{b} and \textbf{d}, while $R_{xy}$ in Fig.~\ref{fig:model}\textbf{e} varies between opposite signs with the same range span.

This MR effect arises from the magnetic field-induced spin precession. It is comparable to the Hanle Magnetoresistance (HMR) in spin-Hall materials~\cite{dyakonov2007magnetoresistance,velez2016hanle,wu2016hanle}, albeit here with CISS as the spin--charge conversion mechanism. We thereby name these effects the CISS-induced Hanle Magnetoresistance (CHMR). It can be distinguished from HMR by its angular dependence. CHMR shows $R_{xx}$ minimum when the field is collinear to the charge current, while HMR shows $R_{xx}$ minimum when the two are perpendicular. This difference is associated with which spin orientations are generated by the charge current through these different mechanisms.

Another important note is that although the CHMR effect can detect whether chirality is present in the 2D film, it cannot determine its sign. This is because both the charge-to-spin and the spin-to-charge conversion processes change sign simultaneously once the chirality (or the sign of CISS) is reversed, and the net result is an unchanged charge signal.

\subsection{Role of spin relaxation}

\begin{figure}[!]
	\includegraphics[width=\linewidth]{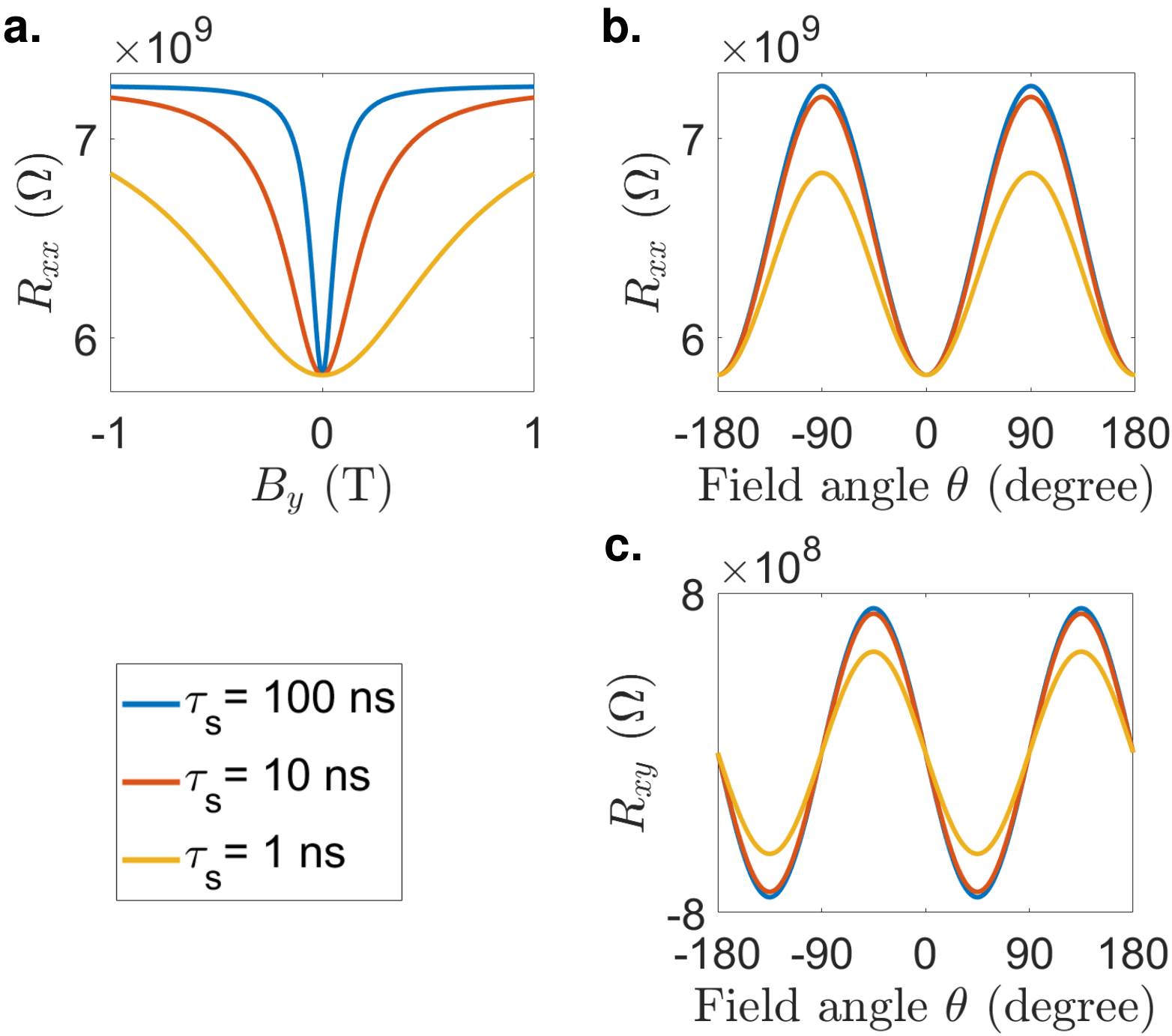}
	\caption{\label{fig:compare}The CISS-induced Hanle Magnetoresistance (CHMR) effect at different spin lifetimes $\tau_s=100$, $10$, and $1$~ns. ($\tau_s=100$~ns was used in Fig.~\ref{fig:model}.) \textbf{a}. The longitudinal sheet resistance $R_{xx}$ as a function of perpendicular magnetic field strength $B_y$ for different spin lifetimes. \textbf{b-c}. $R_{xx}$ and $R_{xy}$ as a function of in-plane magnetic field angle $\theta$ for different spin lifetimes.}
\end{figure}

Another process included in Eq.~\ref{eqn:steady} is spin relaxation (in addition to the spin-flip reflections inherent to CISS), which is characterized by the spin lifetime $\tau_s$. In Fig.~\ref{fig:compare}\textbf{a}, we show how $\tau_s$ affects the dependence of $R_{xx}$ on the perpendicular magnetic field $B_y$. Increasing spin lifetime narrows the Hanle dip around zero field and increases the $R_{xx}$ value at finite field strengths. For low magnetic field strengths that are generally experimentally available, higher $\tau_s$ strengthens the magnetic field dependence and makes the CHMR effect easier to observe.

In Fig.~\ref{fig:compare}\textbf{b-c}, we show how $\tau_s$ affects the angular dependence of the CHMR effect on an in-plane rotating magnetic field. The spin lifetime only affects the amplitude of the sinusoidal angular dependence, i.e. the range in which $R_{xx}$ and $R_{xy}$ are modulated. Increasing spin lifetime increases this range and makes the effect easier to observe.

Here we have compared spin lifetimes from $1$~ns to $100$~ns. Choosing this range is based on the assumed low spin-orbit coupling in most carbon-based organic materials.

\subsection{Multi-terminal device geometries}
\begin{figure}[!]
	\includegraphics[width=\linewidth]{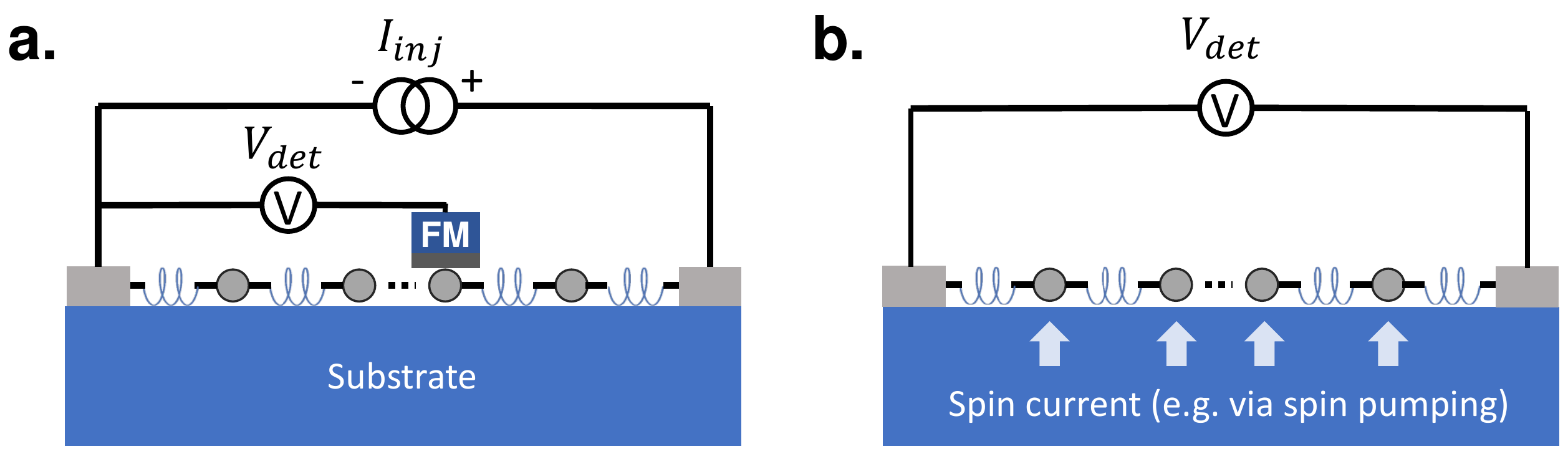}
	\caption{\label{fig:devices}Multi-terminal device geometries for the electrical detection of CISS in linear response. \textbf{a}. Detecting the spin accumulation generated by a charge current via CISS. The outer normal-metal contacts are used to source a charge current, and the inner ferromagnet contact is used for detecting voltage. This voltage is expected to change upon magnetization reversal because of the presence of spin accumulation. \textbf{b}. Detecting the charge voltage generated by a spin current via CISS. The spin current is injected into the chiral film by techniques such as spin pumping. It generates a charge current (hence also a voltage) across the film, which is then detected using normal metal contacts.}
\end{figure}
The above MR effects enable the detection of CISS in the linear response regime, but cannot separately address spin-to-charge and charge-to-spin conversion processes. In Fig.~\ref{fig:devices}, we illustrate two multi-terminal devices that can separate spin and charge contributions, which help to study the \textbf{a}, charge-to-spin and \textbf{b}, spin-to-charge conversion, respectively. 

In Fig.~\ref{fig:devices}\textbf{a}, a charge current $I_{inj}$ is sent through in the chiral film through normal-metal electrodes, and a voltage drop $V_{det}$ can be detected along the current path. If the voltage probe is a ferromagnet, as shown in the figure, an additional voltage can be picked up due to the spin accumulation in the film, which will change upon magnetization reversal. This magnetization-dependent signal allows to evaluate the charge-to-spin conversion property of the chiral film. 

The device in Fig.~\ref{fig:devices}\textbf{b} provides us the access to the reciprocal process, i.e. the spin-to-charge conversion. Here the chiral film is deposited on a surface where spin current can be uniformly injected. This can be done by, for instance, radio-frequency spin pumping from a ferromagnetic insulator~\cite{cornelissen2015long}. The injected spins are converted to a collinear charge current by CISS, which is then detected by the normal-metal voltage probes. 

The voltage signals in both device geometries depend on the spin accumulation in the chiral film. This implies that these multi-terminal measurements are also affected by the spin relaxation and spin precession mechanisms introduced for the CHMR. \textcolor{black}{Unlike for CHMR though, the sign of these signals does depend on the chirality of the material, enabling these devices to be used as chirality detectors.}

\section{Discussion}
In summary, we introduced here two new types of MR signals for electrically detecting CISS in the linear response regime. They are generated by changing the magnitude or orientation of the magnetization or a magnetic field. This is in contrast to the previously reported MR signals based on magnetization or magnetic field reversal, which, according to fundamental theories, should only be possible in the nonlinear response regime.

We first considered the case of a common two-terminal spin valve device, where the new type of MR is generated by changing the net magnetization (magnitude) parallel to the current path. This can be done by rotating the total magnetization using a magnetic field. This type of MR does not require new device geometries and can be tested using existing devices. Note that this mechanism implies that an incomplete magnetization reversal would also give rise to an MR signal in linear response. This is relevant to test experimentally. 

The second type of MR is what we call CISS-induced Hanle Magnetoresistance (CHMR), and it arises from the spin precession dynamics under the presence of an applied noncollinear magnetic field. We illustrated how it can be detected in a 2D chiral film by measuring the longitudinal and transverse sheet resistances as a function of an external magnetic field that varies in strength or direction. This effects depends on the spin lifetime within the chiral film as well as the strength of the CISS effect. Its magnetic field angular dependence resolves the spin orientation (with respect to a charge current) generated by CISS, and distinguishes it from other spin--charge conversion mechanisms. We have restricted our discussion to an in-plane magnetic field in order to focus on spin-related effects. Should an out-of-plane magnetic field be of concern, orbital effects such as Hall effect should also be taken into account. Moreover, other spin dynamics will also contribute to these effects and should be carefully analyzed for specific devices.  

\section{Acknowledgments}
This work is supported by the Zernike Institute for Advanced Materials (ZIAM) and the Spinoza prize awarded to Professor B.~J.~van~Wees by the Nederlandse Organisatie voor Wetenschappelijk Onderzoek (NWO).

\appendix
\section{Appendix}
\subsection{A. Steady-state solution for longitudinal and transverse sheet resistances}  
In the 2D network model, we describe CISS as an averaged property that is evenly distributed within the film. We separately concentrate the spin--charge conversion mechanisms in the chiral elements and (additional) relaxation mechanisms in the nodes. This constructs a repeating pattern in both $x$- and $y$-directions, which is analogous to the repeating unit cells in 2D crystals. When electrically contacted as shown in Fig.~\ref{fig:model}, each unit cell behaves identically at steady state. We can therefore derive the steady-state solution for the longitudinal and transverse conduction through the 2D chiral film. 

\begin{figure}[ht!]
	\includegraphics[width=\linewidth]{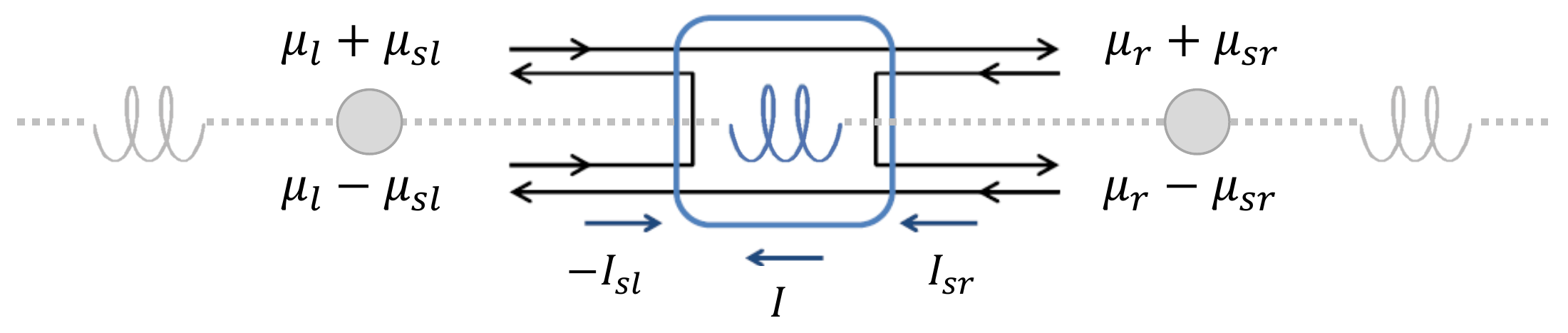}
	\caption{\label{fig:S1}Spin--charge interconversion by a chiral unit.}
\end{figure}

We first look at a single chiral unit along the $x$-direction which is contacted by a node on each side, as illustrated in Fig.~\ref{fig:S1}. The chiral unit generates spin-dependent electron transmission and reflections, which is illustrated using the black arrows in the same way as in Ref.~\cite{yang2020detecting}. The node on the left (right) side of the chiral unit is described by a charge electrochemical potential $\mu_l$ ($\mu_r$) and a spin accumulation $\mu_{sl}$ ($\mu_{sr}$). The spin-specific electrochemical potentials are therefore $\mu_l\pm \mu_{sl}$ ($\mu_r\pm \mu_{sr}$), whose differences drive the spin-specific electron transmission and reflection processes, giving rise to spin currents $I_{sl}$, $I_{sr}$, and charge current $I$. 

Under this description, following the transport matrix formalism introduced in Ref.~\cite{yang2020detecting}, the coupled spin--charge transport equation for this chiral unit and the neighboring nodes is 
\begin{equation}\label{eqn:SCconv}
	\begin{pmatrix}
		I\\-I_{sl}\\I_{sr}
	\end{pmatrix}=-\frac{Ne}{h}\begin{pmatrix}
		t & s & s \\
		P_r r & \gamma_r & \gamma_t \\
		P_t t & \gamma_t & \gamma_r
	\end{pmatrix} \begin{pmatrix}
		\mu_l-\mu_r \\ \mu_{sl} \\ \mu_{sr}
	\end{pmatrix},
\end{equation}
where $N$ is number of spin-degenerate channels, $e$ is elemental charge (positive value), $h$ is Planck's constant, and matrix elements $t$, $s$, $P_t$, $P_r$, $\gamma_t$, and $\gamma_r$ describe the transport property of the chiral unit and is introduced in Ref.~\cite{yang2020detecting}.

When the 2D film is electrically contacted in the way shown in Fig.~\ref{fig:model} and is at steady state, all chiral units (along the $x$ direction) are identical and the electrochemical potential drop across each of them is identically $\Delta\mu_x$. Also, all nodes are identical and they support the same spin accumulation $\mu_s$ and the same CISS-induced spin current injection $I_{s,CISS}$. We therefore have
\begin{align}
	\Delta \mu_x&=\mu_l-\mu_r,\notag \\
	\mu_{sl}&=\mu_{sr}=\mu_s, \notag\\
	I_{s,CISS}&=I_{sr}-I_{sl} \\
	&=-\frac{2Ne}{h}\left( s \Delta \mu_x + (\gamma_r+\gamma_t) \mu_s \right),\notag\\
	\frac{\partial n_{s,CISS}}{\partial t}&=-\frac{I_{s,CISS}}{e}= \alpha \Delta\mu_x +\beta n_s, \notag
\end{align}
where we define $\alpha=2Ns/h$ and $\beta=N(\gamma_r+\gamma_t)/h\nu$ to simplify further derivations. Here we used $\mu_s=n_s/2\nu$ where $\nu$ is the spin-degenerate density of states in the node and $n_s$ is the spin density.

To extend this to a 2D film, we take into account that the charge and spin currents flow in both $x$- and $y$-directions while the spin orientation also includes the $z$-component. The vector equation Eq.~1 can therefore be decomposed into its spacial components
\begin{subequations}
	\begin{equation}
		\widehat{x}: \; -\frac{n_{sx}}{\tau_s}+\gamma(B_y n_{sz}-B_z n_{sy})+\alpha \Delta\mu_x +\beta n_{sx}=0,
	\end{equation}
	\begin{equation}
		\widehat{y}: \; -\frac{n_{sy}}{\tau_s}+\gamma(B_z n_{sx}-B_x n_{sz})+\alpha \Delta\mu_y +\beta n_{sy}=0,
	\end{equation}
	\begin{equation}
		\widehat{z}: \; -\frac{n_{sz}}{\tau_s}+\gamma(B_x n_{sy}-B_y n_{sx})=0,
	\end{equation}
\end{subequations}
where $\gamma=2\mu_B/\hbar$ is the gyromagnetic ratio for electrons in the 2D film, and we assume here the $g$-factor is $2$.

In addition, we have the continuity condition of the charge currents along both $x$- and $y$-directions (derived from Eq.~\ref{eqn:SCconv})
\begin{subequations}
	\begin{equation}
		I_x=-\frac{Ne}{h}(t\Delta \mu_x + 2s \mu_{sx} )=A \Delta \mu_x +B n_{sx},
	\end{equation}
	\begin{equation}
		I_y=-\frac{Ne}{h}(t\Delta \mu_y + 2s \mu_{sy} )=A \Delta \mu_y +B n_{sy}=0,
	\end{equation}
\end{subequations}
where we define $A=-Net/h$ and $B=-Nes/h\nu$ to simplify further derivations, and we used $\mu_s=n_s/2\nu$ to rewrite spin accumulation in terms of spin density. For $I_y=0$ we assume the top and bottom contacts are voltage probes and there is no net charge current entering them.

We consider $n_{sx}$, $n_{sy}$, $n_{sz}$, $\Delta \mu_{x}$, and $\Delta \mu_{y}$ as a function of longitudinal current $I_x$, and rewrite the above five linear equations in matrix form
\begin{equation}
	\begin{pmatrix}
		-\frac{1}{\tau_s}+\beta & -\gamma B_z & \gamma B_y & \alpha & 0 \\
		\gamma B_z & -\frac{1}{\tau_s}+\beta & -\gamma B_x & 0 & \alpha \\
		-\gamma B_y & \gamma B_x & -\frac{1}{\tau_s} & 0 & 0 \\
		B & 0 & 0 & A & 0\\
		0 & B & 0 & 0 & A\\
	\end{pmatrix}
	\begin{pmatrix}
		n_{sx} \\ n_{sy} \\ n_{sz} \\ \Delta\mu_x \\ \Delta \mu_y
	\end{pmatrix}=
	\begin{pmatrix}
		0 \\ 0 \\ 0 \\ I_x \\ 0
	\end{pmatrix}.
\end{equation} 

We name the $5\times 5$ matrix $M$, and the solution for the above equation is
\begin{equation}
	\begin{pmatrix}
		n_{sx} \\ n_{sy} \\ n_{sz} \\ \Delta\mu_x \\ \Delta \mu_y
	\end{pmatrix}= M^{-1}
	\begin{pmatrix}
		0 \\ 0 \\ 0 \\ I_x \\ 0
	\end{pmatrix}
\end{equation}

With this, we establish the relation between $\Delta \mu_x$, $\Delta \mu_y$, and $I_x$. These quantities are evaluated for a unit area only. To calculate the longitudinal and transverse resistances of the film, we need to take into account its length $L$ (along $x$) and width $W$ (along $y$). We assume a square film with $L=W$, and derive the sheet resistances
\begin{subequations}
	\begin{equation}
		R_{xx}=\frac{V_{xx}}{W I_x}=-\frac{L \Delta\mu_x}{e W I_x}=-\frac{\Delta\mu_x}{e I_x},
	\end{equation}
	\begin{equation}
		R_{xy}=\frac{V_{xy}}{W I_x}=-\frac{W \Delta\mu_y}{e W I_x}=-\frac{\Delta\mu_y}{e I_x}.
	\end{equation}
\end{subequations}
These are the quantities plotted in the main text.

Note that we have assumed a uniform film where the electrodes cover entire edges. This may not be the case in practice, and one needs to also consider effects related to the width of the contact, the spin relaxation length in the film, and spin accumulation on sample edges. For the experimental geometry we proposed in Fig.~\ref{fig:devices} where the spin detection with an ferromagnet or spin injection with spin pumping takes place out of the film plane, one should also consider the thickness of the film and how that compares to the spin relaxation length.

\subsection{B. Parameters for plots}
In Fig.~\ref{fig:2TMR}\textbf{b-e}, we tune either the ferromagnet or CISS polarization parameter. We vary one of them between $\pm1$, and keep the other constant at $0.5$. All other parameters are set according to the zero-bias point in Fig.~\ref{fig:model}(d) of Ref.~\cite{yang2020detecting}, and the conductances are calculated following the formalism developed in Ref.~\cite{yang2020detecting}.

For Fig.~\ref{fig:model}, we assume the CISS polarization is $0.5$ for each chiral element and the available spin-degenerate channels through each chiral element is 10. All other parameters for the chiral element are assumed the same as in Fig.~\ref{fig:2TMR}. For the film we assume the density of states to be $10^{15}$~eV$^{-1}$m$^{-2}$. The spin lifetime $\tau_s$ is set as $100$~ns. Figure~\ref{fig:compare} uses the same set of parameters but with different spin lifetimes as specified.
 
\textcolor{black}{\subsection{C. Reciprocity of spin currents}
To further illustrate the Onsager reciprocity, we show here the reciprocal relation between the spin currents driven by spin accumulations in either leads. We consider the same two-terminal circuit as in Fig.~\ref{fig:2TMR}\textbf{a}, but now consider finite spin accumulation in the left and right leads.}

\begin{figure}[ht!]
	\includegraphics[width=\linewidth]{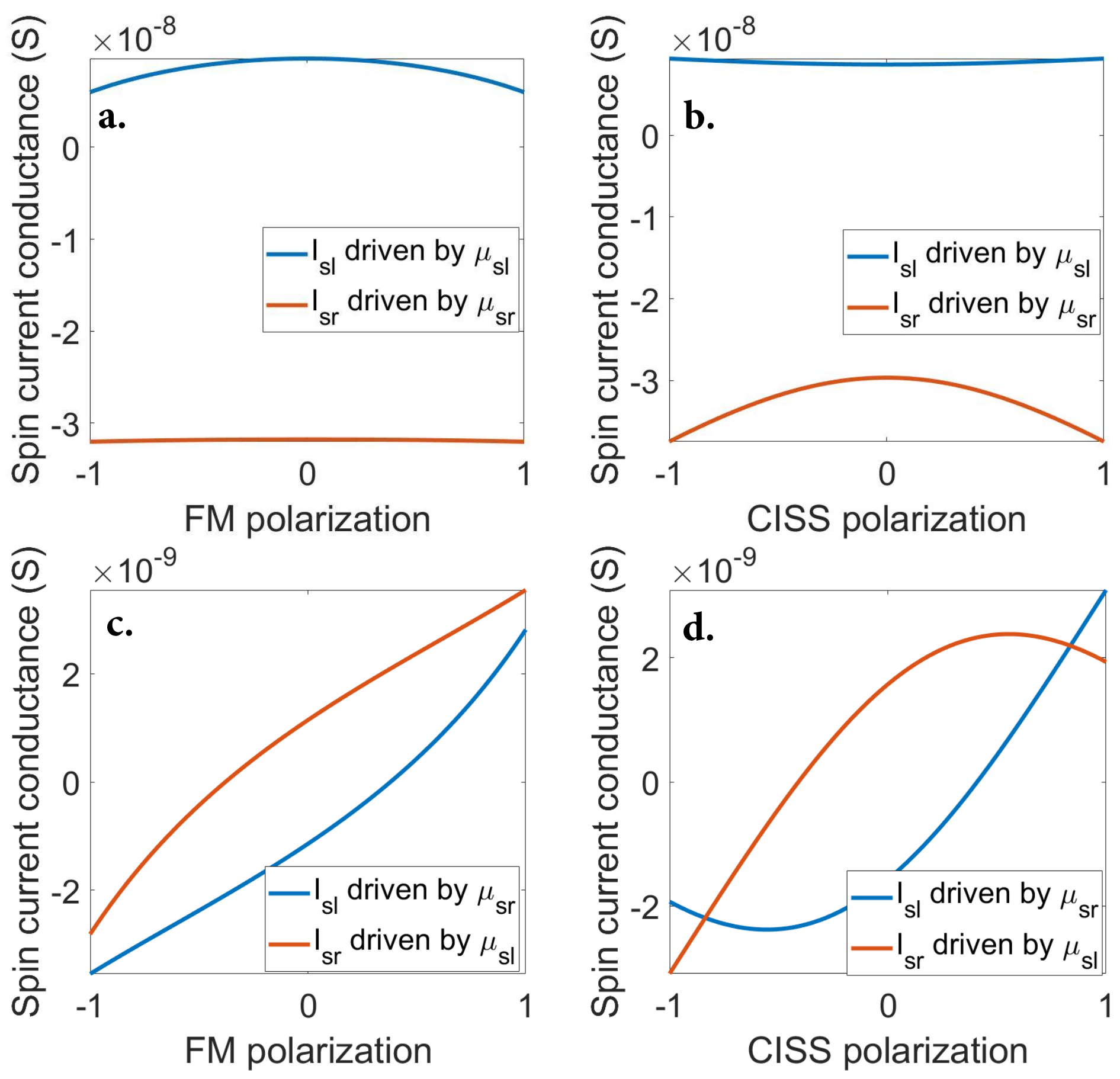}
	\caption{\label{fig:S2}Reciprocity of spin currents in the leads driven by spin accumulations in the same (\textbf{a, b}) and opposite (\textbf{c, d}) lead. The circuit parameters for obtaining these curves are identical as those in Fig.~\ref{fig:2TMR}.}
\end{figure} 

\textcolor{black}{We illustrate here a case where the charge and spin transport is purely driven by spin accumulation. For this, we consider zero bias voltage, i.e. $\mu_L=\mu_R$, and calculate the spin currents injected into one lead per unit spin accumulation in the same/opposite lead. We term this the spin current conductance, and show how it depends on the FM polarization and CISS polarization in Fig.~\ref{fig:S2}. Here $I_{sl}$ ($I_{sr}$) is the spin current in the left (right) lead, and $\mu_{sl}$ ($\mu_{sr}$) is the spin accumulation in the left (right) lead.}

\textcolor{black}{The Onsager reciprocity is highlighted by the symmetries of these curves. In Fig.~\ref{fig:S2}\textbf{a-b}, we show the spin currents in a lead driven by the spin accumulation in the same lead. This spin current is an even function of both FM polarization and CISS polarization. This shows that a full magnetization (or chirality) reversal will not change the spin currents produced by a spin accumulation in the lead itself.}

\textcolor{black}{In Fig.~\ref{fig:S2}\textbf{c-d}, we show how a spin accumulation in one lead drives spin current in the opposite one. Here reciprocity implies specific symmetry relation between the two curves in each panel: they are central symmetric with respect to the origin. Therefore, the two curves in each panel obtain a sign difference when the FM or CISS polarization is reversed for one of them.}

\end{document}